\documentclass[a4paper,11pt]{article}
\usepackage{pos}
\usepackage{lineno}

\title{Comparing HAWC blazars light curves with different data reconstruction versions}
 \ShortTitle{P4 Vs P5 comparison}

\author*[a]{J. A. Garc\'ia-Gonz\'alez}
\affiliation[a]{Tecnologico de Monterrey, Escuela de Ingenier\'ia y Ciencias, Ave. Eugenio Garza Sada 2501, Monterrey, N.L., Mexico, 64849}
\author[]{M. M. Gonz\'alez$^b$ for the HAWC Collaboration}
\affiliation[b]{Instituto de Astronom\'ia, UNAM, Ciudad de M\'exico, Mexico.}

\emailAdd{anteus79@tec.mx}
\emailAdd{magda@astro.unam.mx}

\abstract{We present a comparison of the flux normalization of HAWC sources using 17 months of data that was processed using two different versions of the official data reconstruction used for HAWC analyses. Pass4 (P4) has been used so far for most of the results published by HAWC. The most recent reconstruction version, Pass5 (P5) will be used in future analyses and comes with better pointing accuracy and improved gamma/hadron separation. The aim of this work is to do a comparison of the light curves obtained with both P4 and P5 and show that the results are consistent withing statistical uncertainties. }

\ConferenceLogo{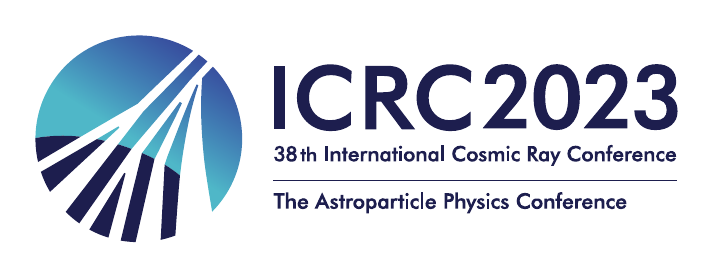}

\FullConference{%
38th International Cosmic Ray Conference (ICRC2023)\\
  26 July - 3 August, 2023\\
  Nagoya, Japan}

\begin{document}
\maketitle

\section{Introduction}\label{intro}
Previously, the HAWC collaboration presented in ~\cite{4} a comparison of the two fitting frameworks used to obtain  light curves (LCs), in particular for three of the main sources being monitored by HAWC: the Crab Nebula, and the blazars Mrk 421 and Mrk 501. The analysis was performed with data in the energy range from $300$ GeV to $100$ TeV using the first 17 months of data. A likelihood method was used to construct daily maps on sidereal days, then a specific spectral model was used to estimate the photon fluxes per every transit for the three sources. In this work we will focus only on Mrk421 and Mrk501.
\section{Official reconstruction pass versions}\label{sec:passv}
HAWC has developed several versions of the reconstruction software. The HAWC observatory has published results using the two software versions, Pass 1~\cite{2} and Pass~4, the latter has been used for the rest of the publications until 2022.

P5 is the current official version and Pass~6 which includes an extended array of outriggers is still under development. In this work we use P4 as reference since it is the latest version that was used in previous analyses of blazars~\cite{1,3} and we will show that there is consistency with P5, this will allow us to extend several analysis that were done using P4.

P5 is a new version of the reconstruction software with better pointing accuracy and improved gamma/hadron separation. A more detailed version of the improvements of P5 can be found at~\cite{6}.
\section{Light Curves}\label{sec:LCs}
For both P4 and P5 reconstruction versions we have fit to the data the spectral model described in Equation~\ref{equ:SED} and the model's parameters are shown in Table~\ref{tab:SEDpar}, these values are consistent with the ones used in~\cite{1}. To obtain the flux normalization $N_{0}$ we use \b{ the Zenith Band Response Analysis (ZEBRA)}~\cite{5}.
\begin{equation}\label{equ:SED}
    \frac{dN}{dE} = N_{0}\left( \frac{E}{E_{0}}\right)^{-\alpha}\exp\left(\frac{E}{E_{c}}\right)
\end{equation}{}
The values for $E_{0}(TeV)$ and $\alpha$ were fixed and the normalization $N_0$ was the only free parameter. The integrated flux $> 1$TeV is obtained later on to calculate the LCs.
\begin{table}[ht]
    \centering
    \begin{tabular}{c|c|c|c}
        source & $E_{0}(TeV)$ & $\alpha$ & $E_{c}(TeV)$ \\
        \hline
        Mrk~421 & 1 & 2.2 & 5\\
        Mrk~501 & 1 & 1.6 & 6
    \end{tabular}
    \caption{Parameters used in the cut-off power law for the observed spectra of Mrk~421 and Mrk~501}
    \label{tab:SEDpar}
\end{table}

Once we have calculated the flux normalization $N_{0}$ we can obtain LCs for the three sources of interest for both P4 and P5. It is important to mention that the days used to calculate the flux are not the same, this is due to quality cuts that removed some days in P5 reconstruction, and there are a few more days where the coverage or other factors can lead to low statistics and a flux can not be obtained. Further investigation will be carried out to try to fix this in a following publication.

\begin{figure*}
\begin{center}
\includegraphics[width=0.95\linewidth]{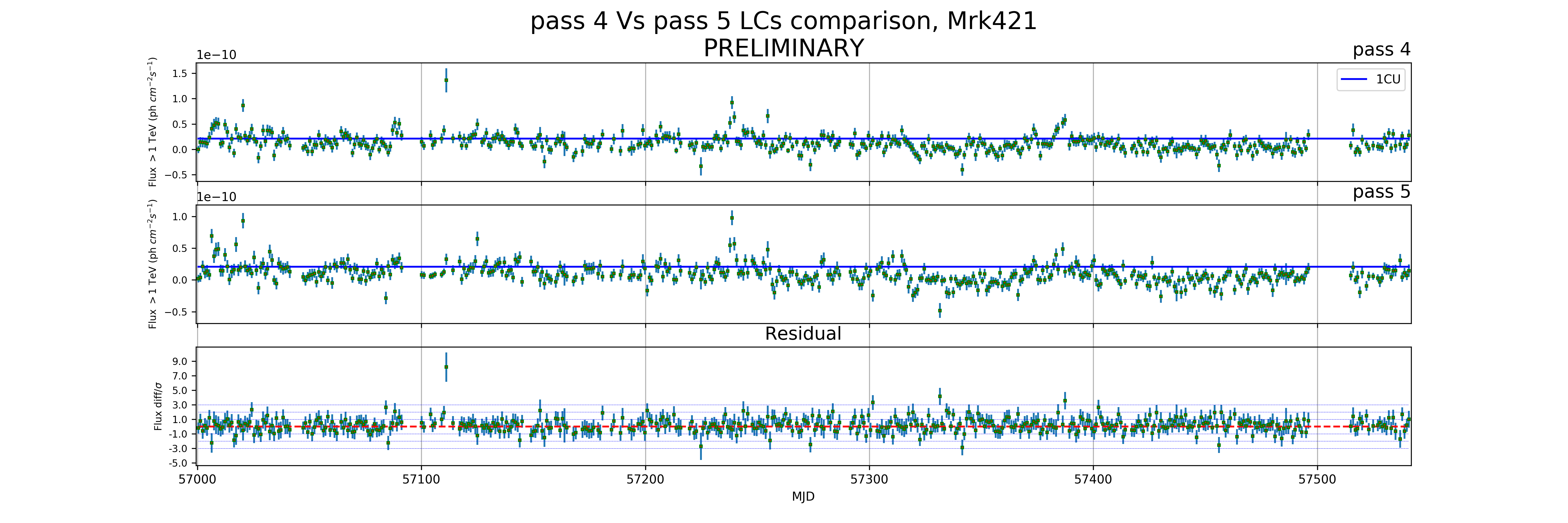}
\caption{LCs comparison for Mrk421. The top plot is P4 data and middle plot is P5 data. Here 17 months of data have been used to compare P4 and P5 reconstruction versions. The blue line in the first two panels represents 1 CU. In the bottom plot a difference between the two upper panels can be found. A discontinuous red line show the zero level in the flux comparison. In the case of Mrk421 there is just one outlier due to a lower flux found in P5.  \label{fig:compMrk421}}
\end{center}
\end{figure*}

\begin{figure*}
\begin{center}
\includegraphics[width=0.95\linewidth]{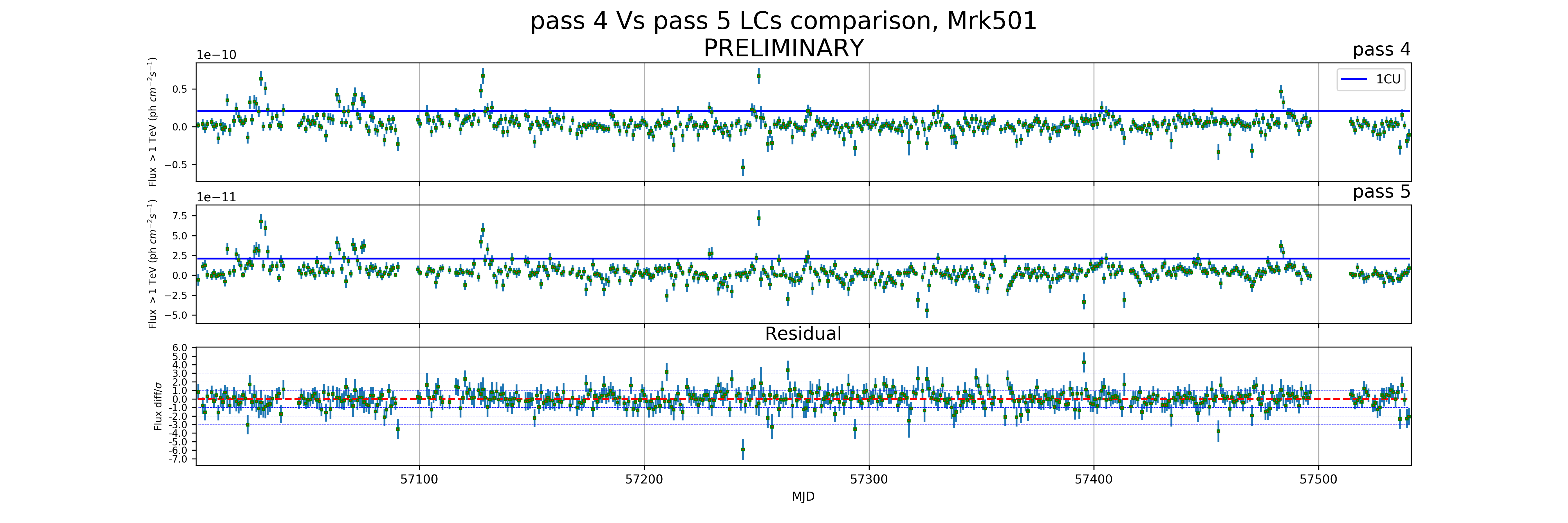}
\caption{LCs comparison for Mrk501.  The top plot is P4 data and middle plot is P5 data. Here 17 months of data have been used to compare P4 and P5 reconstruction versions. The blue line in the first two panels represents 1 CU. In the bottom plot a difference between the two upper panels can be found. A discontinuous red line show the zero level in the flux comparison. In the case of Mrk501 there are a few outliers due to statistical fluctuation for close to zero (or even negative) flux values. \label{fig:compMrk501}}
\end{center}
\end{figure*}

Only the common days will be used for the LCs comparison, yielding to an overlap of  89\% and 93\%  for Mrk421 and Mrk501 respectively. Figure~\ref{fig:compMrk421} and Figure\ref{fig:compMrk501} show the comparison for Mrk~421 and Mrk~501 respectively.

for each plot we have 3 panels, the upper panel is a LC using P4 and the middle panel shows P5. For both plots a blue line is describing a flux level of 1 Crab Unit (CU) as a reference of the overall flux. The lower panel shows the flux difference divided by its standard deviation, which provides an estimate of the dispersion of the data. A couple of blue doted lines to contain $\pm 3\sigma$ are provided to make the comparisons easier for the reader. 

\section{Discussion}
\subsection{Mrk421}
In the case of Mrk421 (Figure~\ref{fig:compMrk421} shows only one outlier due to a high flux present in P4 that is not in P5, the data set used to make the LCs is still preliminary and a further revision will be carried out. For the rest of the LC the two versions are consistent within statistical uncertainties. The gap present around MJD 57500 is due to interruptions in the data, and this will affect any future version of data reconstructed.
\subsection{Mrk501}
For Mrk501 (Figure~\ref{fig:compMrk501}) we have a couple of outliers, in this case very low fluxes cause fluctuations in the results depending on the pass version used, we can see that for the last half of the time range chosen, Mrk501 has exhibit a very low flux, and this is reflected in negative fluxes due to an underestimation of the background. Except for the outliers we can see that both LCs versions are consistent within statistical uncertainties.

\section{Conclusions }\label{sec:conclusions}
We have shown that the new data reconstruction version P5 gives consistent results when been compared with the previous data reconstruction version P4 within statistical uncertainties. We have compared the LCs for both Mrk421 and Mrk501 and the overall LCs show a consistency between P4 and P5 results. There are still a couple of days that show no consistency but considering the time range used, consisting of approximately 518 days for both sources, the two fluxes that need investigation are a minor fix. Further processing of the data sample used for P5 will be carried out in a separate work.

\section*{Acknowlegments}
We acknowledge the support from: the US National Science Foundation (NSF); the US Department of Energy Office of High-Energy Physics; the Laboratory Directed Research and Development (LDRD) program of Los Alamos National Laboratory; Consejo Nacional de Ciencia y Tecnolog\'{i}a (CONACyT), M\'{e}xico, grants 271051, 232656, 260378, 179588, 254964, 258865, 243290, 132197, A1-S-46288, A1-S-22784, CF-2023-I-645, c\'{a}tedras 873, 1563, 341, 323, Red HAWC, M\'{e}xico; DGAPA-UNAM grants IG101323, IN111716-3, IN111419, IA102019, IN106521, IN110621, IN110521 , IN102223; VIEP-BUAP; PIFI 2012, 2013, PROFOCIE 2014, 2015; the University of Wisconsin Alumni Research Foundation; the Institute of Geophysics, Planetary Physics, and Signatures at Los Alamos National Laboratory; Polish Science Centre grant, DEC-2017/27/B/ST9/02272; Coordinaci\'{o}n de la Investigaci\'{o}n Cient\'{i}fica de la Universidad Michoacana; Royal Society - Newton Advanced Fellowship 180385; Generalitat Valenciana, grant CIDEGENT/2018/034; The Program Management Unit for Human Resources \& Institutional Development, Research and Innovation, NXPO (grant number B16F630069); Coordinaci\'{o}n General Acad\'{e}mica e Innovaci\'{o}n (CGAI-UdeG), PRODEP-SEP UDG-CA-499; Institute of Cosmic Ray Research (ICRR), University of Tokyo. H.F. acknowledges support by NASA under award number 80GSFC21M0002. We also acknowledge the significant contributions over many years of Stefan Westerhoff, Gaurang Yodh and Arnulfo Zepeda Dominguez, all deceased members of the HAWC collaboration. Thanks to Scott Delay, Luciano D\'{i}az and Eduardo Murrieta for technical support.

\bibliographystyle{plain}



%
%
%
\clearpage
\section*{Full Authors List: \ Collaboration}
\noindent \textbf{Note comment afterwards:} Collaborations have the possibility to provide an authors list in xml format which will be used while generating the DOI entries making the full authors list searchable in databases like Inspire HEP. \\
\scriptsize
\noindent
%
\vskip2cm
\noindent
A. Albert$^{1}$,
R. Alfaro$^{2}$,
C. Alvarez$^{3}$,
A. Andrés$^{4}$,
J.C. Arteaga-Velázquez$^{5}$,
D. Avila Rojas$^{2}$,
H.A. Ayala Solares$^{6}$,
R. Babu$^{7}$,
E. Belmont-Moreno$^{2}$,
K.S. Caballero-Mora$^{3}$,
T. Capistrán$^{4}$,
S. Yun-Cárcamo$^{8}$,
A. Carramiñana$^{9}$,
F. Carreón$^{4}$,
U. Cotti$^{5}$,
J. Cotzomi$^{26}$,
S. Coutiño de León$^{10}$,
E. De la Fuente$^{11}$,
D. Depaoli$^{12}$,
C. de León$^{5}$,
R. Diaz Hernandez$^{9}$,
J.C. Díaz-Vélez$^{11}$,
B.L. Dingus$^{1}$,
M. Durocher$^{1}$,
M.A. DuVernois$^{10}$,
K. Engel$^{8}$,
C. Espinoza$^{2}$,
K.L. Fan$^{8}$,
K. Fang$^{10}$,
N.I. Fraija$^{4}$,
J.A. García-González$^{13}$,
F. Garfias$^{4}$,
H. Goksu$^{12}$,
M.M. González$^{4}$,
J.A. Goodman$^{8}$,
S. Groetsch$^{7}$,
J.P. Harding$^{1}$,
S. Hernandez$^{2}$,
I. Herzog$^{14}$,
J. Hinton$^{12}$,
D. Huang$^{7}$,
F. Hueyotl-Zahuantitla$^{3}$,
P. Hüntemeyer$^{7}$,
A. Iriarte$^{4}$,
V. Joshi$^{28}$,
S. Kaufmann$^{15}$,
D. Kieda$^{16}$,
A. Lara$^{17}$,
J. Lee$^{18}$,
W.H. Lee$^{4}$,
H. León Vargas$^{2}$,
J. Linnemann$^{14}$,
A.L. Longinotti$^{4}$,
G. Luis-Raya$^{15}$,
K. Malone$^{19}$,
J. Martínez-Castro$^{20}$,
J.A.J. Matthews$^{21}$,
P. Miranda-Romagnoli$^{22}$,
J. Montes$^{4}$,
J.A. Morales-Soto$^{5}$,
M. Mostafá$^{6}$,
L. Nellen$^{23}$,
M.U. Nisa$^{14}$,
R. Noriega-Papaqui$^{22}$,
L. Olivera-Nieto$^{12}$,
N. Omodei$^{24}$,
Y. Pérez Araujo$^{4}$,
E.G. Pérez-Pérez$^{15}$,
A. Pratts$^{2}$,
C.D. Rho$^{25}$,
D. Rosa-Gonzalez$^{9}$,
E. Ruiz-Velasco$^{12}$,
H. Salazar$^{26}$,
D. Salazar-Gallegos$^{14}$,
A. Sandoval$^{2}$,
M. Schneider$^{8}$,
G. Schwefer$^{12}$,
J. Serna-Franco$^{2}$,
A.J. Smith$^{8}$,
Y. Son$^{18}$,
R.W. Springer$^{16}$,
O.~Tibolla$^{15}$,
K. Tollefson$^{14}$,
I. Torres$^{9}$,
R. Torres-Escobedo$^{27}$,
R. Turner$^{7}$,
F. Ureña-Mena$^{9}$,
E. Varela$^{26}$,
L. Villaseñor$^{26}$,
X. Wang$^{7}$,
I.J. Watson$^{18}$,
F. Werner$^{12}$,
K.~Whitaker$^{6}$,
E. Willox$^{8}$,
H. Wu$^{10}$,
H. Zhou$^{27}$

\vskip2cm
\noindent
$^{1}$Physics Division, Los Alamos National Laboratory, Los Alamos, NM, USA,
$^{2}$Instituto de Física, Universidad Nacional Autónoma de México, Ciudad de México, México,
$^{3}$Universidad Autónoma de Chiapas, Tuxtla Gutiérrez, Chiapas, México,
$^{4}$Instituto de Astronomía, Universidad Nacional Autónoma de México, Ciudad de México, México,
$^{5}$Instituto de Física y Matemáticas, Universidad Michoacana de San Nicolás de Hidalgo, Morelia, Michoacán, México,
$^{6}$Department of Physics, Pennsylvania State University, University Park, PA, USA,
$^{7}$Department of Physics, Michigan Technological University, Houghton, MI, USA,
$^{8}$Department of Physics, University of Maryland, College Park, MD, USA,
$^{9}$Instituto Nacional de Astrofísica, Óptica y Electrónica, Tonantzintla, Puebla, México,
$^{10}$Department of Physics, University of Wisconsin-Madison, Madison, WI, USA,
$^{11}$CUCEI, CUCEA, Universidad de Guadalajara, Guadalajara, Jalisco, México,
$^{12}$Max-Planck Institute for Nuclear Physics, Heidelberg, Germany,
$^{13}$Tecnologico de Monterrey, Escuela de Ingeniería y Ciencias, Ave. Eugenio Garza Sada 2501, Monterrey, N.L., 64849, México,
$^{14}$Department of Physics and Astronomy, Michigan State University, East Lansing, MI, USA,
$^{15}$Universidad Politécnica de Pachuca, Pachuca, Hgo, México,
$^{16}$Department of Physics and Astronomy, University of Utah, Salt Lake City, UT, USA,
$^{17}$Instituto de Geofísica, Universidad Nacional Autónoma de México, Ciudad de México, México,
$^{18}$University of Seoul, Seoul, Rep. of Korea,
$^{19}$Space Science and Applications Group, Los Alamos National Laboratory, Los Alamos, NM USA
$^{20}$Centro de Investigación en Computación, Instituto Politécnico Nacional, Ciudad de México, México,
$^{21}$Department of Physics and Astronomy, University of New Mexico, Albuquerque, NM, USA,
$^{22}$Universidad Autónoma del Estado de Hidalgo, Pachuca, Hgo., México,
$^{23}$Instituto de Ciencias Nucleares, Universidad Nacional Autónoma de México, Ciudad de México, México,
$^{24}$Stanford University, Stanford, CA, USA,
$^{25}$Department of Physics, Sungkyunkwan University, Suwon, South Korea,
$^{26}$Facultad de Ciencias Físico Matemáticas, Benemérita Universidad Autónoma de Puebla, Puebla, México,
$^{27}$Tsung-Dao Lee Institute and School of Physics and Astronomy, Shanghai Jiao Tong University, Shanghai, China,
$^{28}$Erlangen Centre for Astroparticle Physics, Friedrich Alexander Universität, Erlangen, BY, Germany

\end{document}